\documentstyle[11pt,aaspp4]{article}

\def \kms {{\rm km~s^{-1}}}
\def \kpc {{\rm kpc}}

\def \Msun {{{\rm M}_{\sun}}}
\def \yr {{\rm yr}}

\def \Md {{\dot{M}}}

\def \baryon {{{\rm baryon}}}
\def \BH {{{\rm BH}}}
\def \blowout {{{\rm blowout}}}
\def \bulge {{{\rm bulge}}}
\def \circular {{{\rm circ}}}
\def \collapse {{{\rm collapse}}}
\def \cool {{{\rm cool}}}

\def \crit {{{\rm crit}}}
\def \disk {{{\rm disk}}}
\def \DM {{{\rm DM}}}
\def \dyn {{{\rm dyn}}}
\def \edd {{{\rm edd}}}
\def \esc {{{\rm esc}}}
\def \gas {{{\rm gas}}}
\def \H {{{\rm H}}}
\def \halo {{{\rm halo}}}

\def \init {{{\rm init}}}
\def \Kelvin {{{^\circ}{\rm K}}}
\def \max {{{\rm max}}}
\def \shock {{{\rm shock}}}
\def \SNI {{{\rm SNI}}}
\def \SNII {{{\rm SNII}}}
\def \spin {{{\rm spin}}}
\def \universe {{{\rm universe}}}
\def \virial {{{\rm virial}}}

\def \BDM {\begin{displaymath}}
\def \EDM {\end{displaymath}}
\def \BEQ {\begin{equation}}
\def \EEQ {\end{equation}}
\def \BEQA {\begin{eqnarray}}
\def \EEQA {\end{eqnarray}}
\def \NN {\nonumber}
\def \BL {\begin{list}}
\def \EL {\end{list}}
\def \BENUM {\begin{enumerate}}
\def \EENUM {\end{enumerate}}
\def \BITEM {\begin{itemize}}
\def \EITEM {\end{itemize}}
\def \BARR {\begin{array}}
\def \EARR {\end{array}}
\def \BFIG {\begin{figure}}
\def \EFIG {\end{figure}}

\lefthead{KEPNER}
\righthead{Inside-out Galaxy Formation}

\begin{document}

\title{
Inside-out Galaxy Formation}


\author{
  Jeremy~V.~Kepner
}
\affil{
Princeton University Observatory \\
Peyton Hall, Ivy Lane, Princeton, NJ 08544--1001 \\
jvkepner@astro.princeton.edu \\
Current address: MIT Lincoln Laboratory, Lexington, MA}

\begin{abstract}

  Current theories of galaxy formation have tended to focus on
hierarchical structure formation, which is the most likely scenario for
cosmological models with lots of power at small scales (e.g. standard
cold dark matter). Models with little small scale power lead to
scenarios closer to spherical collapse.  Recently favored power spectra
(e.g. CDM+$\Lambda$) lie somewhere in between suggesting that both types
of processes are important and may vary over time due to gaseous
reheating.  From this viewpoint this paper explores a very simple inside
out scenario for galaxy formation. This scenario is a natural result of
synthesizing earlier work on DM halos, spherical collapse, and gas
redistribution via angular momentum. Although, this model is highly
simplified and is not designed to accurately describe the detailed
formation of any individual galaxy, it does (by design) predict the
overall features of galaxies.  In addition, old bulges and young disks
are an almost unavoidable  result of these very simple models.  This
scenario may provide a useful framework for both observers and
theoreticians to think about galaxy formation.

\end{abstract}



\section{Introduction}

  Galaxy formation theories can be caricatured by two models:
hierarchical clustering and spherical collapse; with reality most likely
lying somewhere in between. The degree of hierarchical clustering is
strongly dependent upon upon the amount of power on small scales.
Cosmological models with lots of small scale power (e.g., standard CDM)
will be more hierarchical, while many of the cosmological models
currently under consideration have power spectra with less small scale
power (e.g., mixed dark matter, tilted CDM+$\Lambda$) (\cite{Borgani97})
and will be less hierarchical.

  The thermal history of the gas can also play an important role.  At
early epochs the gas temperature will be low, which allows small lumps
to form and be accreted hierarchically.  At later epochs, the IGM heats
up to $\sim10^4~\Kelvin$ due to the first generation of stars and Quasar
emission.  Gas at this temperature will be driven out of halos with
circular velocities less than $30~\kms$ (\cite{Kepner97}).  Accretion of
these small lumps at $z \sim 3$ may not take place hierarchically but
will tend towards spherical infall as the puffed up lumps of gas
encounter gas in the halo that has been heated by shocks and radiation
from a variety of sources: stars, supernova, and possibly a supermassive
black hole.  In reality, the gas in galaxy formation is most likely
in multiple phases (like the present day galaxy).

  Exploring both hierarchical and spherical collapse aspects of galaxy
formation is best accomplished with a variety of techniques. N-body
simulations are best for exploring hierarchical structure formation.
Detailed simulations of the formation of individual galaxies can now be
performed with an ever increasing list of physical processes. Some of
the important results of simulations are the approximate correctness of
Press-Schechter theory for describing merging histories
(\cite{Lacey94}), that dark matter (DM) halos have a similar form over a
wide range of scales (\cite{Navarro96}), and the nature of the
Ly$\alpha$ forest (\cite{Miralda-Escude96}, \cite{Ma97},
\cite{Zhang97}).  In addition, simulations that include gas dynamics
show development of disks as a result of tidal torques
(\cite{Navarro97}, \cite{Steinmetz95}). 

  Analytic methods lend themselves to the understanding of spherical
collapse.  The original analytic work on spherical collapse
(\cite{Gunn72}, \cite{Fillmore84}, \cite{Bertschinger85}) laid the
groundwork for understanding virialization, self-similar collapse, the
establishment of near isothermal dark matter halos, and secondary
infall.  More recent analytic work has been particularly useful in
providing intuition on the formation of disk galaxies
(\cite{Dalcanton97}, \cite{Mo97}), an area that is particularly
challenging for N-body simulations due to the large dynamic range
requirements.  These recent analytic papers, built upon the earlier work
of \cite{Mestel63}, \cite{Fall80}, and \cite{Gunn82}, indicates that
allowing gas with typical angular momentum distributions to settle into
typical DM halos produces disk galaxies that reproduce a wide variety of
observations from faint surface brightness objects to damped Ly$\alpha$
systems.  The fundamental assumptions of \cite{Dalcanton97} and
\cite{Mo97}, which will also be adopted in this paper, are that (1)
there is little angular momentum transport, and (2) the angular momentum
eventually halts the collapse of the gas, resulting in a roughly flat
rotation curve.

  Spherically symmetric analytic models have a limited ability to
address highly non-linear phenomena such as hydrodynamic shocks.  Recent
spherical numerical simulations of infall into galaxies indicate that
gas can undergo a shock before settling into the DM potential
(\cite{Thoul95}).  Such shocks are potentially an important process for
slowing down and heating infalling gas during galaxy formation. 

  The goal of this paper is to add to the analytical understanding of
galaxy formation by connecting the earlier spherical collapse models
with the more recent work on redistribution by angular momentum and
infall shocks.  The approach is quite simple.  Given a known approximate
initial state---some kind of primordial ``bump'' (\cite{Bardeen86}), and
a known approximate final state of the DM halo and the gaseous disk,
completing the picture requires determining the time, mass and length
scales for the transformation of the DM and the gas from the initial
state to the final state.  The collapse and virialization of spherically
symmetric DM halos gives an estimate of the scales involved in creating
a DM halo.  The scales for gaseous collapse can likewise be set by the
evolution of an outwardly moving shock (\cite{Thoul95}). 

 Synthesizing these two components results in a single physically
motivated scenario which, by construction, reproduces many of the
observable quantities of the final states of galaxies, but also makes
specific predications about gas evolution during galaxy formation. Such
a scenario can be a useful perspective for both observers and
theoreticians by clarifying the important quantities to be measured and
the additional physical processes that need to be included to explain
various phenomena.

  In \S 2 a sketch of the inside-out galaxy formation scenario is
presented along with an overview of the underlying physics.  In \S 3 the
physical model is presented in detail. \S 4 discusses the results of
these calculations. \S 5 gives conclusions and plans for further work.

\section{Inside-out Galaxy Formation}

  Any exploration of galaxy formation begins with a list of ingredients. 
The two main components are dark matter and gas.  In this paper it is
assumed that the dark matter is cold and collisionless.  The gas can
behave in a variety of ways depending upon its thermal history.  If the
gas is dense it can radiate away heat acquired in shocks and will tend
to behave in a relatively cold manner.  If the gas is diffuse it may be
heated by external radiation or shocks.  Recent simulations of infalling
gas with radiative cooling indicates that such shocks occur in a variety
of galactic potentials (\cite{Thoul95}).  Furthermore, these simulations
suggest that the initial radius at which a mass shell is shocked is
similar in both the adiabatic and non-adiabatic cases.  Thus the
simplest approach is to use a adiabatic numerical approach to calculate
the initial shock radius and then use analytic non-adiabatic methods to
estimate the post-shock behavior.  In reality, the gas in galaxy
formation is most likely in multiple phases (like the present day
galaxy).  Of course, the limits of the adiabatic approximation for
computing the shock radius must be kept in mind in interpreting any
subsequent results. 

  The inside-out scenario that results from combining spherical collapse
and angular momentum redistribution proceeds as follows.  Initially the
gas and dark matter are coupled and expand with the Hubble flow.  The
innermost shells turn around first.  The dark matter virializes, and the
gas is shock heated to the virial temperature.  The cooling time for
these inner shells is very short and some of the gas may quickly
condense into cold lumps and form bulge stars.  Later, the outermost
shells turn around and their gas is shock heated.  However, for these
outer shells, the higher virial temperatures and lower gas densities
result in longer cooling times; so that as the gas cools and falls to
its corresponding angular momentum radius it is shocked again and forms
stars.  Higher angular momentum material ends up outside the bulge and
forms the disk.  Lower angular momentum material falls into the bulge,
which continues to grow slowly, but may be halted by energy input from
supernova and/or a massive black hole that can initiate a wind that
blows out the gas. 

  The two components of the above scenario are DM and gas.  In the next
two sub-sections the specific evolution of each component is elaborated.
In addition,  some of the concepts that form the basis of the physical
model will also be introduced.

\subsection{DM Evolution}

  Typically, a spherical dark matter halo is quantified by a
characteristic mass and radius ($M_\halo$,$r_\halo$), which can be
translated into a circular velocity and virialization redshift
($V_\circular$,$z_\virial$) by assuming a value for the overdensity at
$z_\virial$ (e.g., $\delta_\virial \sim 179$).  Simple estimates of the
number of halos of a given mass at each redshift can be made via the
Press-Schecter formalism.  These estimates provide useful insights into
the size distribution of DM halos and are in approximate agreement with
galaxy catalogs and N-body simulations. 

  As primordial DM halos initially expand and contract they are torqued
by neighboring density peaks leaving some galaxies with significant
amounts of angular momentum.  The amount of angular momentum is
characterized by the spin parameter $\lambda_\spin$.  Galaxies with
larger values of $\lambda_\spin$ form disks.  Thus three parameters
would seem to describe the gross characteristics of galaxies
($M_\halo$,$r_\halo$,$\lambda_\spin$) or equivalently
($V_\circular$,$z_\virial$,$\lambda_\spin$).

  For an idealized spherically symmetric DM halo, the evolution is
qualitatively quite simple.  Each mass shell in a primordial ``bump''
proceeds through Hubble expansion, turn around, virialization and
relaxation into a final quasi-stationary DM profile.  This process is
characterized by the collapse time $t_\collapse$ and the maximum
radius $r_\max$ of each bound DM shell.

\begin{figure}
\plotone{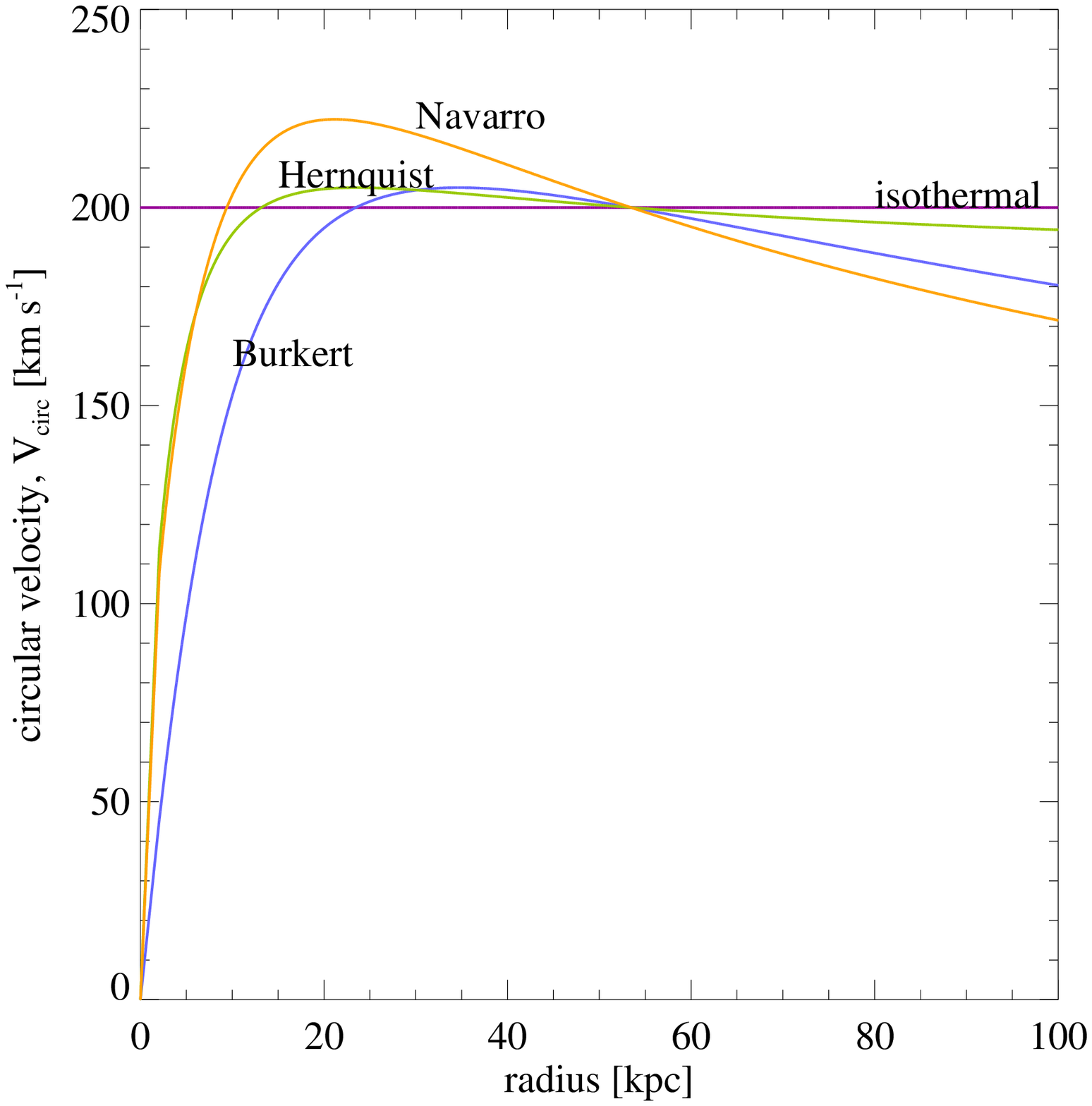}
\caption{
\label{rotation_curves}
  The circular velocities for different profiles for a
DM halo with $V_\circular(r_\halo) = 200~\kms$, $z_\virial = 3$, $M_\halo =
10^{12}~\Msun$, $r_\halo = 54~\kpc$.  All the profiles reproduce typical
rotation curves over the observable range. Hernquist: $M \propto r^2/(r + a)^2$,
isothermal: $M \propto r$, Navarro: $M \propto \log(r^2 + a^2)$,
Burkert: $\rho \propto [(r + a)(r^2 + a^2)]^{-1}$, where $5 a = r_\halo$.
} \end{figure}

  The initial bump can be estimated via various theoretical arguments
and the result can usually be fit by an approximately Gaussian profile. 
The final DM profiles have been extensively studied both observationally
and via N-body simulations and a variety of universal profiles have been
proposed.  These differ in detail but produce similar rotation curves,
$V_\circular(r)$, over the observationally accessible radii (see
Figure~\ref{rotation_curves}).  The process of virialization has also
been extensively studied in simulations.  There is no simple way to
characterize this process.  Fortunately, for spherical collapse this
process seems to take place fairly quickly and in a manner which tends
to erase the details of the initial distribution (\cite{Hoffman88}). 

  An asymptotically correct model of the DM evolution can be constructed
by marrying the initial expansion and turn around of a pre-determined
initial profile with a pre-determined final profile by simply stopping
each infalling mass shell at the radius set by the final profile. Such a
model---while completely sidestepping the important role of
virialization and hierarchical structure formation---provides a simple
framework for describing the DM potential and its effect on the gas.

\subsection{Gaseous Evolution}
  For the most part, each gas shell of the galaxy evolves passively
under the gravitational influence of the dark matter until it collapses. 
In the case of adiabatic gas, the inner shells will be compressed until
there is sufficient pressure to stop the infall and create a shock. 
This process is characterized by the shock time $t_\shock$ and shock
radius $r_\shock$ of each mass shell which define the pertinent time and
length scales of the gas in the halo.  Inside the shock, the gas will be
approximately in isothermal hydrostatic equilibrium with a temperature
given by the DM potential at $r_\shock$.  Outside the shock, the gas
will be cooler and significantly less dense.  In a real multi-phase gas,
$r_\shock$ can be thought of as an approximate upper limit for the
radius at which cooler clumps of gas begin to feel the effects of the
external environment.   The ``true'' radius will be somewhat smaller
to account for the fraction of gas that has condensed into cold lumps.
Some simulations indicate that early on this can be a large fraction
of the gas (\cite{Navarro97}).  Later on the ratio of cold
gas to shocked gas can be as low as 20\% (\cite{Thoul95}), suggesting
that the adiabatic value of $r_\shock$ becomes more accurate at later
times.

  The main effects of the shock are twofold.  First, the dissipation of
a significant amount of inward velocity results in the slow spiraling of
any cool clumps towards their corresponding angular momentum radii. 
Second, a large increase in density leads to a dramatic drop in the
cooling time $t_\cool \propto \rho_\gas^{-1}$.  If the cooling time is
short compared to the dynamical time $t_\dyn = \pi r V_\circular^{-1}$,
then even the clumps in the hot gas will condense and may undergo rapid
star formation, otherwise they will spiral inwards until they hit the
disk.  The initial gas mass and radius of this region, which may
correspond to an initial bulge, can be computed by setting $t_\dyn =
t_\cool$.  Subsequent mass shells will split their material between the
bulge region and the disk depending upon the distribution of specific
angular momentum within each mass shell.  The bulge will continue to
grow slowly.  However, this process can be halted if there is sufficient
energy input from supernovae and/or a massive black hole to drive a
wind.  The disk will continue to grow outward as long as there is
infalling material to supply it. 

  Asymptotically redistributing the gas according to its angular
momentum produces disks in agreement with a wide variety of observations
(\cite{Dalcanton97}, \cite{Mo97}).  Thus, as with the DM halo, there is a
fairly good picture of the early and late stages of the gas.  The
evolving adiabatic shock is a means for setting the time, length and mass
scales of the evolution of the gas between these states.

\section{Physical Model}

  The previous section gave a broad overview of an inside-out scenario
for galaxy formation and hinted at the underlying physical model used to
motivate it.  In this section the details are presented. First, the
final DM halo is described.  Second, the evolution of the DM halo from a
primordial bump is discussed.  Third, the evolution of the shock radius
of the gas is presented.  Fourth, the initial formation of the bulge is
computed. Fifth, the infall of material onto the bulge and the disk is
derived.  Sixth, the processes that may lead to a wind are discussed.

\subsection{Final DM halo}
  Let the final dark matter halo be specified by two parameters: the halo
radius and halo mass, which can be translated into the the circular
velocity $V_\circular$ and the virialization redshift $z_\virial$ if it is
assumed that the  overdensity at virialization is $\delta_\virial \sim
179$ (\cite{Gunn72})
  \BEQ
       V_\circular^2(r_\halo) = \frac{G M_\halo}{r_\halo} ~ , ~~~ 
      \frac{4\pi}{3} r_\halo^3 \delta_\virial \rho_\crit(z_\virial) = M_\halo ,
  \EEQ
where the mean density is given by the usual expressions for a
$\Omega=1$ CDM cosmological model: $\rho_\crit(z) = (1 + z)^3
\rho_\crit^0$, $6 \pi G \rho_\crit t_\universe^2 = 1$, $t_\universe^0 = 2/3
H_0$.   The distribution within the DM halo can be any reasonable
profile (e.g. \cite{Hernquist90}, \cite{Burkert95}, \cite{Navarro96}, see
Figure~\ref{rotation_curves}), which satisfies the above equations at
$r_\halo$.  The simplest choice is an isothermal sphere for which:
  \BEQ
       M_\DM(r) = \frac{M_\halo}{r_\halo} r ~ , ~~~
       V_\circular(r) = V_\circular(r_\halo) ~ , ~~~
       \rho_\DM(r) = \frac{M_\halo}{r_\halo} \frac{1}{4 \pi r^2}
  \EEQ

\subsection{Primordial bump}
\begin{figure}
\plotone{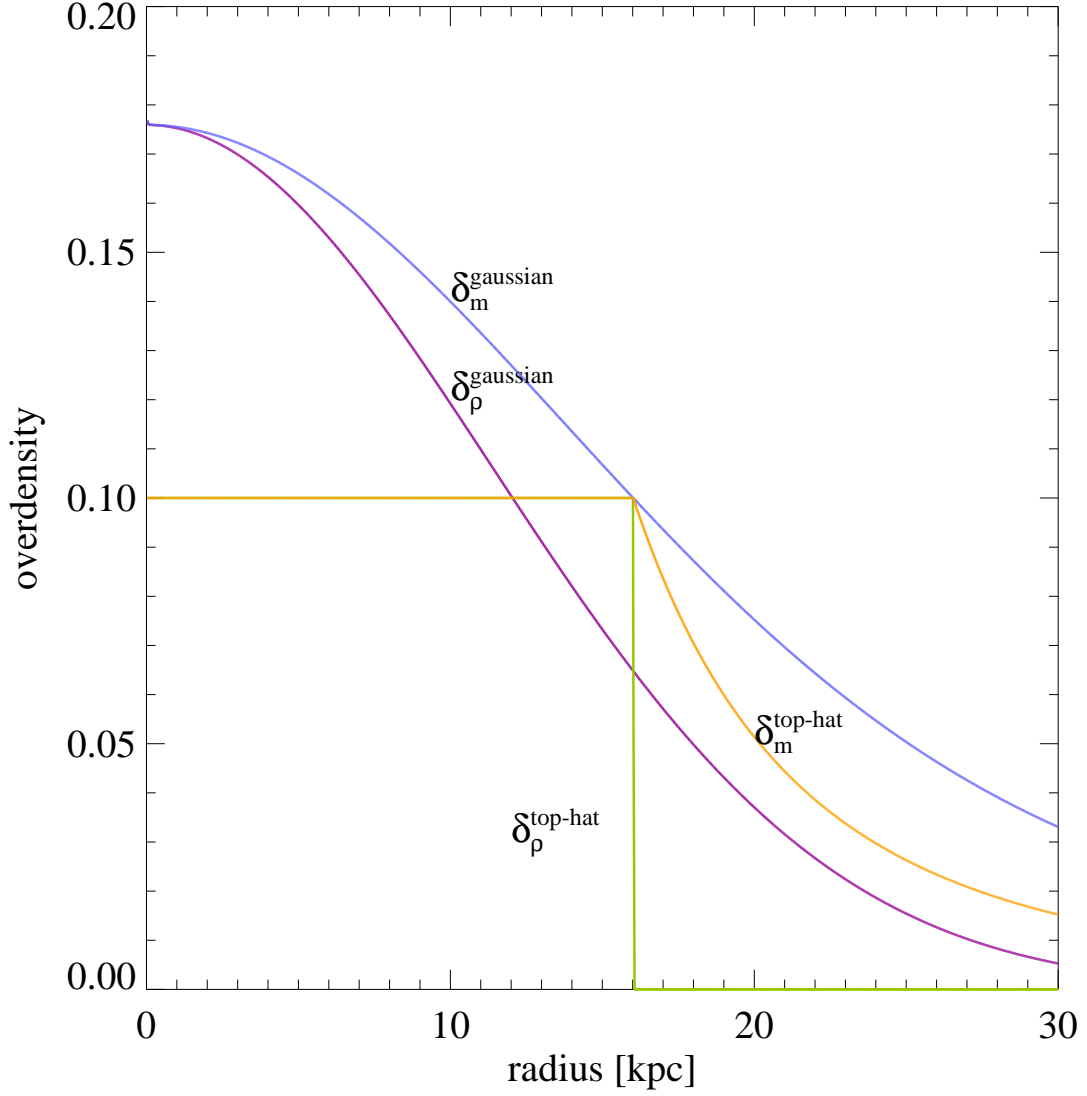}
\caption{
\label{initial_profile}
  Initial density profile and initial mass overdensity profile obtained
from a Gaussian bump, and a corresponding top-hat profile for the halo
given in Figure~1 ($z_\init = 68$, $r_\halo(z_\init) = 16~\kpc$).
  } \end{figure}

  The simplest description of the state of a DM halo at some
early epoch (corresponding to redshift $z_\init$)
is a spherical ``bump'' in the density field $\rho_\DM(r,z_\init)$.
A typical initial density profile might be (see Figure~\ref{initial_profile}):
  \BEQ
       \rho_\DM(r,z_\init) = \rho_\crit(z_\init) + \rho_0 e^{-r^2/r_\halo^2(z_\init)}
  \EEQ
The dynamics of each mass shell in the halo are governed by
the interior overdensity:
  \BEQ
       \delta_\DM(r) + 1 = \frac{M_\DM(r)}{
          \frac{4 \pi}{3} r^3 \rho_\crit(z_\init) }
  \EEQ
where $M_\DM(r) =  \int_0^r 4 \pi r'^2 \rho_\DM(r') dr'$.
$\delta_\DM$ determines the Hubble flow for each point
at $z_\init$ via the linear theory estimate (\cite{Thoul95})
  \BEQ
       v_\DM(r,z_\init) = r H(z_\init) [ 1 - \frac{1}{3} \delta_\DM(r)] .
  \EEQ
The above estimate will be accurate while $\delta_\DM << 1$.

  Having specified the initial position, velocity and time
of each mass shell the subsequent evolution is governed
by
  \BEQ
       2 \pi \frac{t}{t_\collapse} = \theta - \sin \theta ~ , ~~~
       \frac{2 r}{r_\max} = 1 - \cos \theta ~ , ~~~
       v_\DM = \frac{\pi \sin \theta}{1 - \cos \theta} \frac{r_\max}{t_\collapse},
  \EEQ
where
  \BEQ
       \frac{1}{r_\max(r)} = \frac{1}{r} - \frac{v(r)}{2 G M_\DM(r)} ~ , ~~~
       t_\collapse(r) = 2 \pi \sqrt{\frac{r_\max^3}{8 G M(r)}}
  \EEQ

\begin{figure}
\plotone{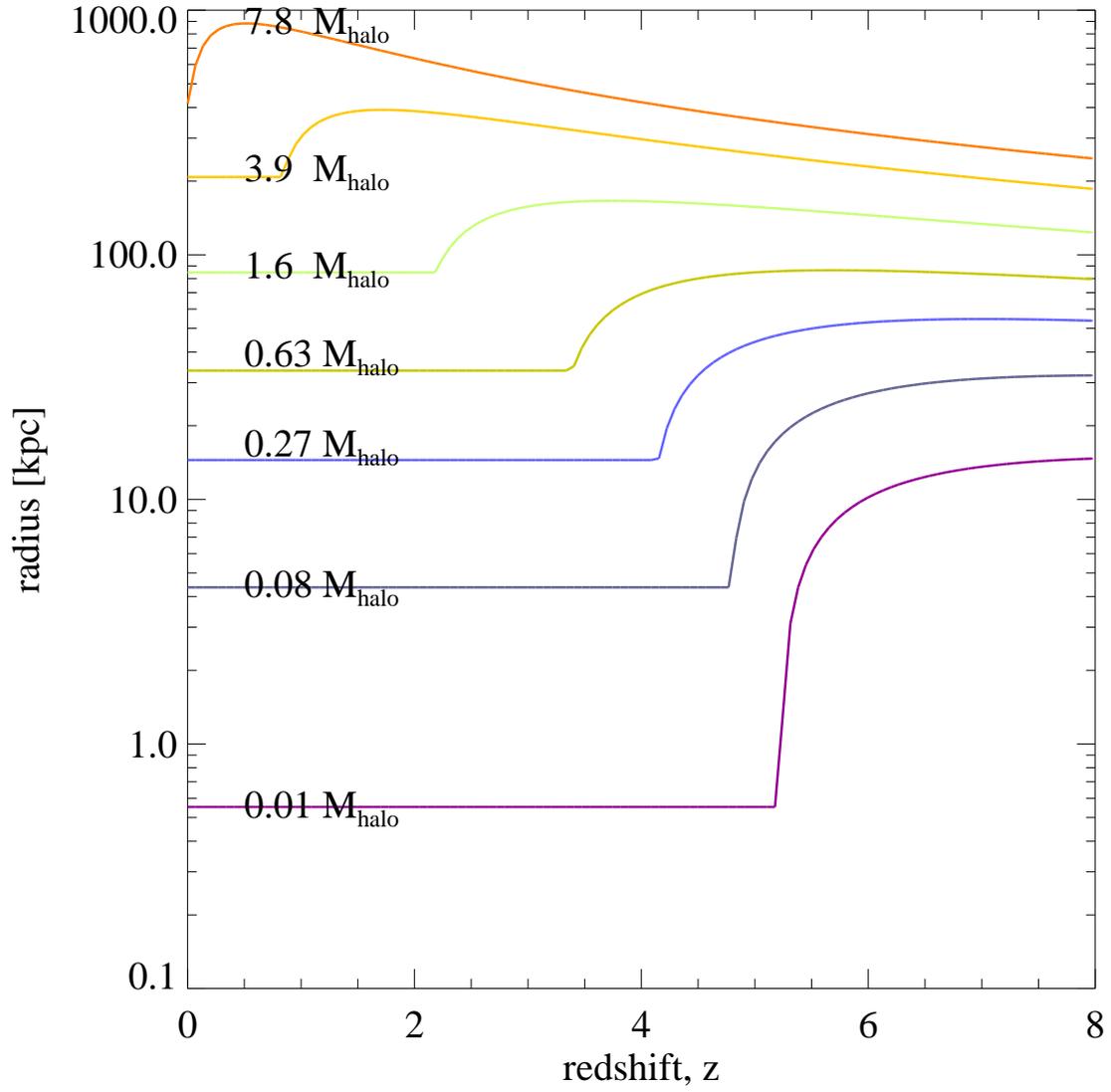}
\caption{
\label{trajectories}
  Trajectories of selected DM shells.  The shells are stopped at the radius
corresponding to the final position for the isothermal halo given in Figure~1.
  } \end{figure}

  The above description of the trajectories holds as long as there are
no shell crossings.  Unfortunately, there is no elegant procedure to
ensure collapsing shells virialize into a desired quasi-stationary DM
halo.  The simplest method, which is adopted here, is to stop each mass
shell when it reaches a pre-determined final radius. It remains to be
seen whether or not this approach provides a sufficiently accurate
picture of the DM potential from the point of view of the gas.
Figure~\ref{trajectories} shows the trajectories of various mass shells
when this method is employed with an isothermal profile.

\subsection{Shock Radius}
  Detailed spherical collapse calculations (\cite{Thoul95};
\cite{Thoul96}) indicate that gas falling into a variety of galaxy
potentials undergoes a shock.  These simulations were conducted for both
adiabatic and non-adiabatic gas (using standard cooling functions) in a
fully dynamical DM potential.  One important observation of these
simulations is that the initial shock radius in both adiabatic and
non-adiabatic gas is similar.  Howver, after the initial shock the gas
cannot be treated adiabatically.  While a non-adiabatic model in a fully
dynamic DM potential provides a richer description of galaxy formation
it is more difficult to develop a simple relations for the shock radius
from such simulations.  This would suggest that the simplest means for
obtaining an expression of the initial shock radius is to examine
adiabatic gas falling into a deterministically evolving DM potential. 
After obtaining this relation, a non-adiabatic analytic approach is
adopted in the next sub-section to look at the post-shock behaviour of
the gas. 

\begin{figure}
\plotone{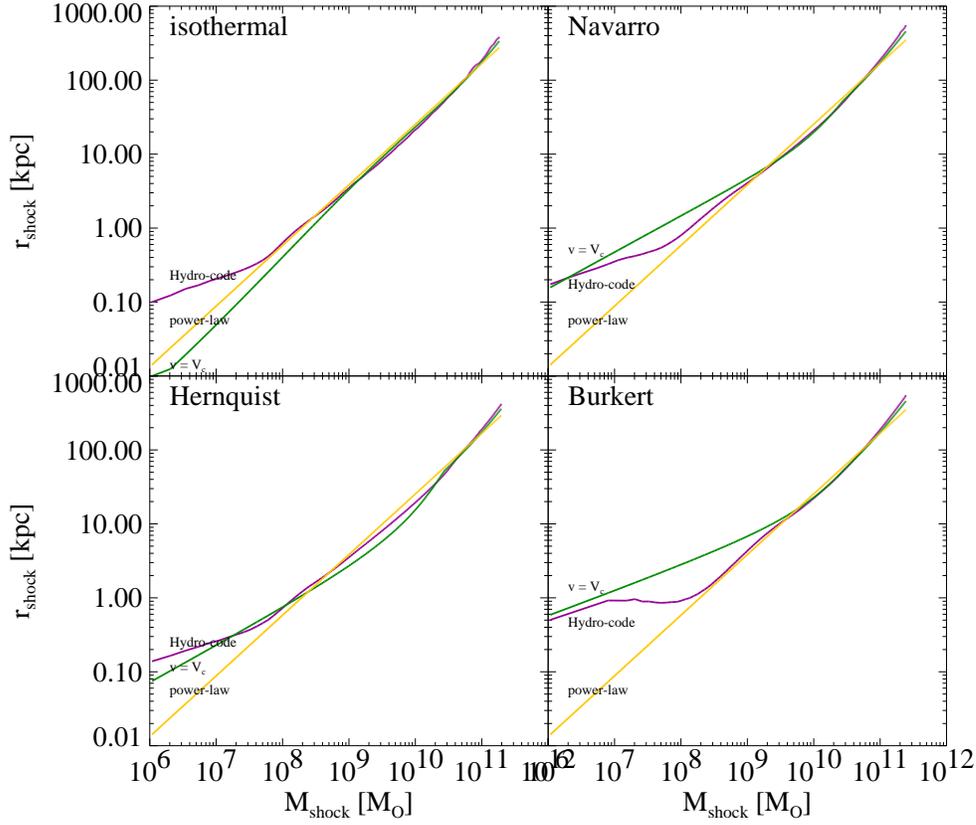}
\caption{
\label{shock_radius}
  Shock radius $r_\shock$ vs shock mass $M_\shock$ for the four DM halos given
in Figure~1 using three methods: (i) Lagrangian hydrodynamics code (ii)
infall velocity constraint $V_\circular(r_\shock) = v_\DM(M_\shock)$ (iii)
simple power law $(r_\shock/r_\halo) = (M_\shock/\Omega_\baryon
M_\halo)^{\beta_\shock}$, ${\beta_\shock} = 0.82$.
  } \end{figure}

  The adiabatic evolution of the gas within the above evolving
DM potential is governed by the Lagrangian equations:
  \BEQA
     \frac{d}{dr} M_\gas &=& 4 \pi r^2 \rho_\gas ~, \NN \\
     \frac{d}{dt} v_\gas &=& - 4 \pi r^2 \frac{d P_\gas}{d M_\gas}
                         - \frac{G M_\DM(r)}{r^2} ~, \NN \\
     \frac{d}{dt} u_\gas &=& \frac{P_\gas}{\rho_\gas^2} \frac{d}{dt} \rho_\gas ~, \NN \\
                  P_\gas &=& (\gamma - 1) \rho_\gas u_\gas ~,
  \EEQA
where the self-gravity of the gas has been ignored because it
becomes important only in the inner regions of the bulge and the disk,
well inside the shock. For the relavent initial conditions
($\rho_\gas(r) = \Omega_\baryon \rho_\DM(r)$, $v_\gas(r) = v_\DM(r)$),
the global behavior of the solutions to the above equations are quite
similar: the gas expands and contracts with the DM until the pressure
gradient stops the infall and causes a shock to form.  Inside the shock
the gas is hot and dense, while outside the shock it is cool and less
dense.  Thus, the evolution of the shock front describes, to first
order, the overall behavior of the gas.  The shock occurs when the infall
speed of a shell approximately equals the thermal speed of
the gas interior to it.  If the gas is assumed to follow the DM until
this point, and that inside the shock the gas is at a temperature
corresponding to the circular velocity of the DM halo, then the shock
will occur when:
  \BEQ
       V_\circular(r_\shock) = v_\gas(M_\shock) \approx v_\DM(M_\shock/\Omega_\baryon,t_\shock),
  \EEQ
where $M_\shock = M_\gas(r_\shock) = \Omega_\baryon M_\DM(r_\shock)$. 
Figure~\ref{shock_radius} shows plots of $r_\shock$ vs.  $M_\shock$ for
the four halos shown in Figure~\ref{rotation_curves} obtained from
solving the above equation and compares it to that obtained using a
spherically symmetric Lagrangian hydro-dynamics code similar to that of
\cite{Thoul95}.  Both results obey a simple power law
  \BEQ
       \frac{r_\shock}{r_\halo} \approx
          \left ( \frac{M_\shock}{\Omega_\baryon M_\halo} \right )^{\beta_\shock}
         ~ ,~~~ {\beta_\shock} \approx 0.82 .
  \EEQ
In the next two sections this result is used to estimate the gas
properties of the bulge and the disk.

\subsection{Initial Bulge}

\begin{figure}
\plotone{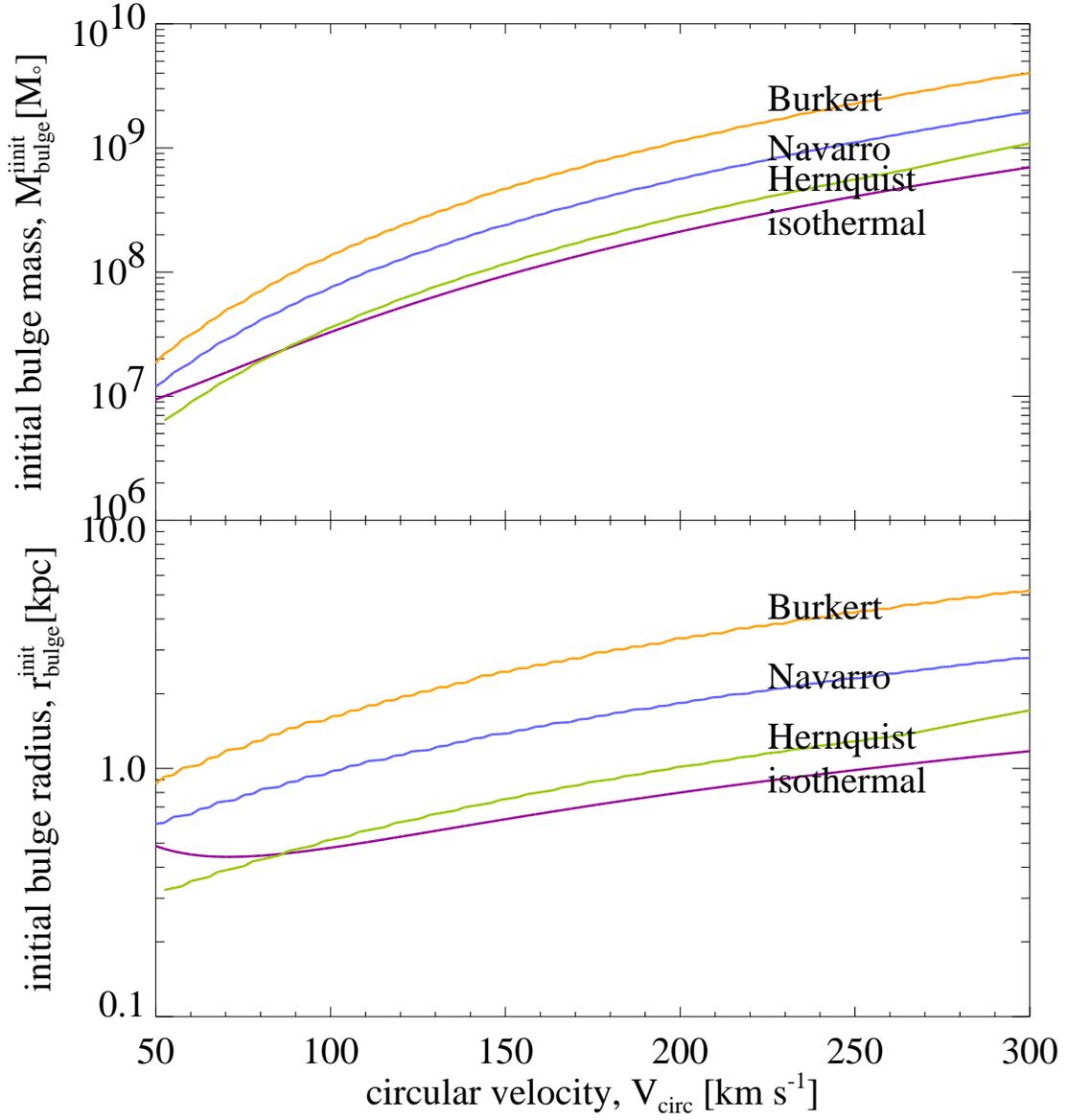}
\caption{
\label{initial_bulge}
  Initial bulge radius $r_\bulge$ and initial bulge mass $M_\bulge$ as a
function of circular velocity computed for the four halos shown in
Figure~1 assuming $z_\virial = 3$. 
  } \end{figure}

  Inside the shock the gas is approximately isothermal with a temperature
given by $T_\shock = \alpha_T V_\circular^2(r_\shock)$.  In addition, the gas
will be in hydrostatic equilibrium.  In the case of a DM halo with
a flat rotation curve, the density of the gas just inside the shock
is
  \BEQ
        \rho_\shock = \rho_\gas(r_\shock) \approx
           \frac{M_\shock}{\frac{4 \pi}{3 - \alpha_T} r_\shock^3} ~ ,
  \EEQ
If the cooling time is short compared to the dynamical time, then clumps
in the shocked gas will condense and form the first bulge stars,
otherwise they will spiral inwards until they hit the disk. The boundary
between gas that cools quickly and gas that hits the disk can be
computed by setting $t_\dyn(r_\shock) = t_\cool(r_\shock)$:
  \BEQ
       \frac{\pi}{2} \frac{r_\shock}{V_\circular}
       = \frac{\rho_\shock k_B T_\shock}{(\gamma-1) \mu}
         \frac{\mu^2}{\Lambda(T_\shock) \rho_\shock^2} ~ ,
  \EEQ
where $\Lambda(T)$ is the primordial optically thin cooling function
(\cite{Thoul95}), $\mu$ is the mean molecular weight, $k_B$ is
Boltzman's constant and $\gamma = \frac{5}{3}$. Manipulating the above
expression gives
  \BEQ
        \frac{M_\shock}{r_\shock^2}
        = \frac{8}{3 - \alpha_T}
          \frac{\mu k_B T_\shock V_\circular}{(\gamma-1) \Lambda(T_\shock)}
  \EEQ
All the terms on the right can be obtained directly from the parameters
specifying the DM halo.  If $M_\shock$ and $r_\shock$ are related by a
simple power law, then the initial mass and radius of the bulge can be
estimated.  Figure~\ref{initial_bulge} shows $r_\bulge^\init$ and
$M_\bulge^\init$ for four DM halos as a function of $V_\circular$
for $\alpha_T = 2$.

\subsection{Disk Formation}
  Subsequent mass shells will split their material between the bulge and
the disk depending upon the distribution of specific angular momentum
within each mass shell.  The simplest model for the initial angular
momentum distribution is to assume that each shell is in solid body
rotation with frequency $\omega_0$ when $r = r_\max$, where $\omega_0 =
\lambda_\spin V_\circular /r_\halo$.  This model is consistent with the
idea that a typical proto-galaxy acquires the majority of its angular
momentum from an single encounter with a similar sized object when it is
at maximum expansion.  In the bump initial conditions used here the
inner shells will turn around before the outer shells.  If the external
torquing field changes in time, then the shells will have different
angular momentum histories.  This is most likely the case in the real
galaxies.  However, because solid body rotation does an adequate job of
explaining the current state of disk galaxies (\cite{Dalcanton97},
\cite{Mo97}), it is convenient to maintain this simplifying assumption. 

  The distribution of specific angular momentum within a rotating shell
is not the same.  Gas at the equator will have more angular momentum
than gas at the poles.  This distribution of specific angular momentum
within a gas shell at a time when the shell is at a radius $r_\gas$ is $j_\gas =
(R_\gas/r_\gas)^2 j_\max$, where $j_\max = \omega_0 r_\max^2$ and
$R_\gas \le r_\gas$ is the projected radius.  The specific angular
momentum of each radius in the DM halo is $j_\DM = V_\circular(R_\DM)
R_\DM$.  The final distribution of the gas can be computed by setting
$j_\gas = j_\DM$
  \BEQ
       M_\gas(R_\DM) = \left \{
                         1 - \sqrt{1 - j_\DM(R_\DM)/j_\max}
                       \right \}  M_\gas^{tot},
  \EEQ
The majority of the mass of each shell ends up near the maximum radius
$j_\DM(R_\DM) = j_\max$, which for an isothermal halo is $R_\DM =
j_\max/V_\circular$.  The final distribution of the disk can be computed by
summing the contributions of each gas shell.  Detailed calculations of
this type reproduce a wide range of observed properties
(\cite{Dalcanton97}, \cite{Mo97}).

  The gas takes approximately $t_\dyn(r_\shock)$ to settle into the
above distribution after it has been shocked.  Because the majority of
each gas shell falls near the maximum angular momentum radius, the disk
grows outward according to $r_\disk^{\max}(t_\shock(M_\shock) +
t_\dyn(M_\shock)) = j_\max(M_\shock)/V_\circular$.  This outer edge will
grow as long as there is infalling material.

\begin{figure}
\plotone{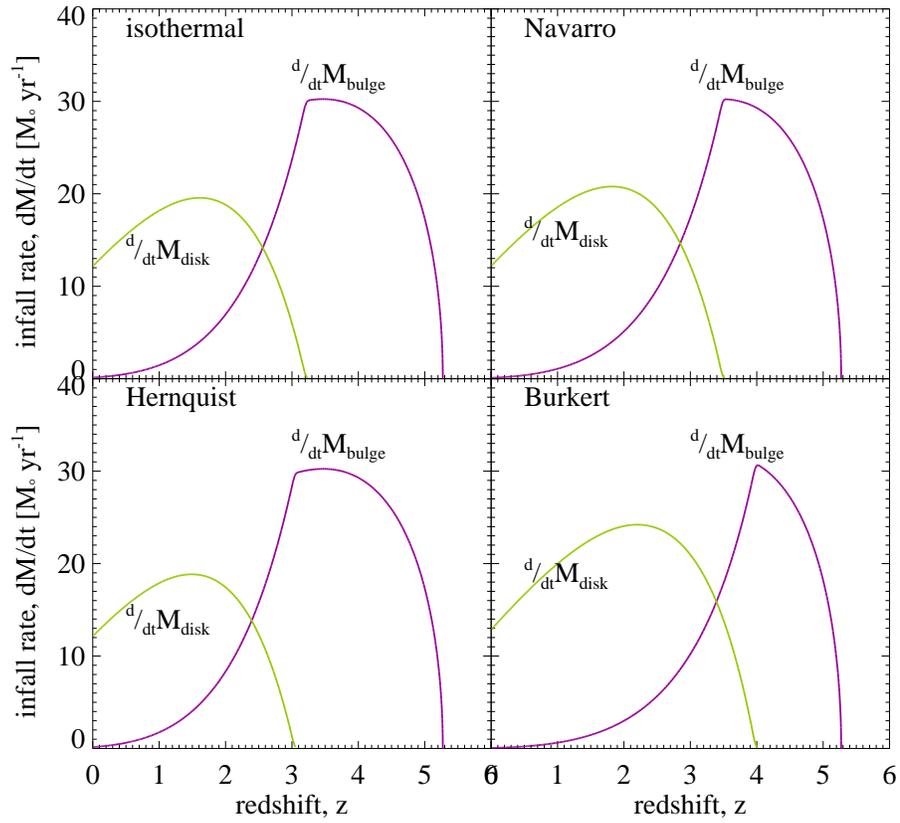}
\caption{
\label{infall}
  Disk and bulge infall rates vs. redshift  for four DM halos
with $V_\circular = 200~\kms$ and $z_\virial = 3$ and $r_\bulge = 4~\kpc$,
$\lambda_\spin = 0.03$.
  } \end{figure}

  In contrast, the bulge continues to grow by slowly accumulating the
low specific angular momentum material of many gas shells.  The total
amount of material falling into the bulge can computed by integrating
the shocked shells
  \BEQ
       M_\bulge(t_\shock(M_\shock) + t_\dyn(M_\shock))
       = \int_0^{M_\shock} \left \{
           1 - \sqrt{1 - j_\DM(r_\bulge)/j_\max(M_\shock')}
         \right \} dM_\shock'
  \EEQ
In general, $t_\dyn + t_\shock \approx t_\collapse$, so for any halo:
  \BEQ
       M_\bulge(t_\collapse(M_\shock))
       = \int_0^{M_\shock} \left \{
           1 - \sqrt{1 - \frac{V_\circular(r_\bulge) r_\bulge r_\halo}
                   {V_\circular(r_\halo) \lambda_\spin r_\max(M_\shock')}}
         \right \} dM_\shock'
  \EEQ
The gas in the disk is simply $M_\disk = M_\shock - M_\bulge$. 
Figure~\ref{infall} shows the evolution of the infall rates $\Md_\bulge$
and $\Md_\disk$ as a function of redshift for four different DM halos. 
Interestingly, this very simple picture produces old bulges and young
disks quite naturally, with reasonable values for the total mass and
infall rates. 

\subsection{Blowout}
  The bulge will continue to grow slowly without additional sources of
energy to heat the gas and drive a wind.  There are three potential
sources of energy input: Supernovae Type I, Supernovae Type II and a
massive black hole. A realistic treatment of the feedback due to any of
these processes is a challenging undertaking and is fundamentally
limited by an incomplete understanding of several processes (e.g., star
formation, supernova IMF, black hole formation, etc ...).  Here only
the most simple estimates are attempted. Fundamentally, to drive
a wind the energy input into the bulge gas must overcome the escape
velocity of the infalling gas:
  \BEQ
       \Md_\bulge V_\esc^2
       < \max\{\alpha_\SNI L_\SNI,\alpha_\SNII L_\SNII,\alpha_\SNI L_\BH\}
  \EEQ
where $V_\esc$ is the typical escape velocity of gas in the bulge, and
$\alpha_{\SNI,\SNII,\BH}$ is the fraction of the total energy input of
each type which is usefully deposited into the gas in such a way as to
drive a wind, i.e.  a parameterization of our ignorance about converting
luminosities into winds.  The energy inputs can be estimated as follows
  \BEQA
         L_\SNI & = & \epsilon_\SNI M_\bulge V_\SNI^2 / t_\SNI \\
         L_\SNII & = & \epsilon_\SNII \Md_\bulge V_\SNII^2 \\
         L_\BH & = & \epsilon_\BH \Md_\BH c^2  \approx L_\edd = M_\BH c^2/t_\edd
  \EEQA
where , $t_\edd = \frac{\sigma_T c}{4 \pi G m_\H} \approx 5 \times
10^9~\yr$.

  The equation for $L_\SNI$ assumes the progenitors of Type I SN (really Type
Ia) are white dwarfs or some other compact stellar remnant which should
be proportional to the total bulge population.  If the
average rate of SNI is one per hundred years per $10^{10}~\Msun$
(\cite{Pain96}) and each SNI ejects approximately $1~\Msun$ of material,
then setting $\epsilon_\SNI \sim 0.01$, $V_\SNI \sim 10^4~\kms$ and
$t_\SNI \sim 10^{10}~\yr$ will give approximately the correct rates and
energies for SNI. 

  The equation for $L_\SNII$ assumes the progenitors of Type II SN
(really Type Ib,c and Type II) are massive short lived stars whose rate
should be proportional to the star formation rate, which in turn is roughly
proportional to the infall rate.  If the average rate of SNII is once
per hundred years (\cite{Cappellaro97}) in galaxies with star formation
rates and infall rates of on the order of one solar mass per year, and
each SNII also ejects  $\sim 1~\Msun$ of material, then setting
$\epsilon_\SNII \sim 0.01$ and $V_\SNI \sim 10^4~\kms$ will give
approximately the correct rates and energies for SNII. 

  The equation for $L_\BH$ assumes the massive black hole radiates
a fraction of the gravitational energy of the material falling onto
it.  Furthermore, it assumes that the luminosity and subsequently the
infall rate is constrained by the Eddington luminosity of the black hole.

  Using these definitions it is possible to try and guess
under what conditions each type of energy input will drive a wind.
The blowout condition for Type II Supernova is the simplest
  \BEQ
         V_\esc < V_\SNII \alpha_\SNII^{\frac{1}{2}}
                          \epsilon_\SNII^{\frac{1}{2}} ~ ,
  \EEQ
The value of $\alpha_\SNII$ is difficult to estimate, but can be
obtained by setting $V_\esc \sim 100~\kms$ the equivalent critical value
arrived at in more detailed calculations (\cite{Dekel86}), in which case
$\alpha_\SNII \sim 0.01$. 

\begin{figure}
\plotone{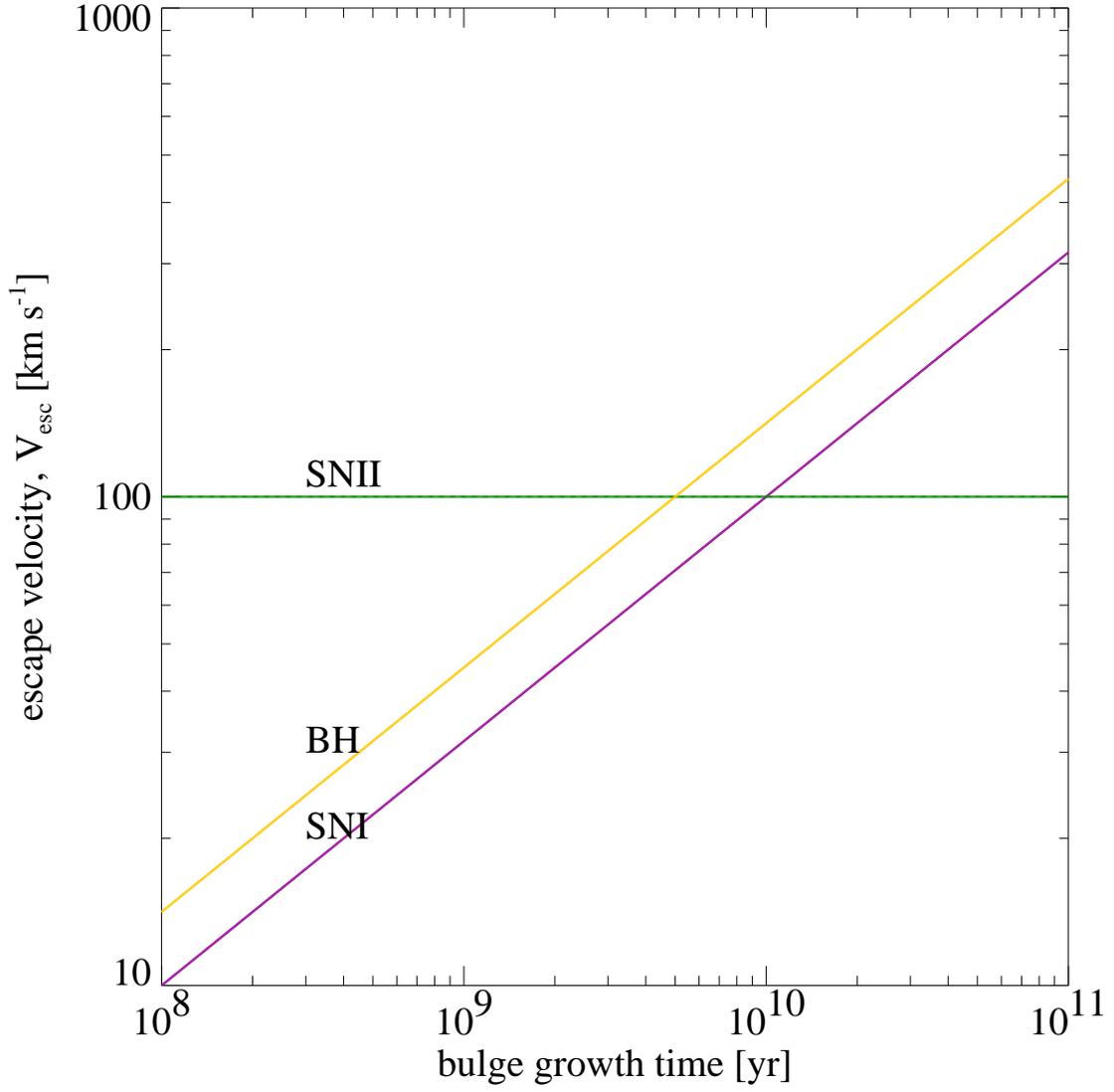}
\caption{
\label{blowout}
  Bulge blowout conditions for Type I and Type II supernova and
a super-massive black hole.
} \end{figure}

  For Type I Supernova the blowout condition is
  \BEQ
      \Md_\bulge V_\esc^2 < \alpha_\SNI \epsilon_\SNI M_\bulge V_\SNI^2 / t_\SNI ~,
  \EEQ
which can be rewritten in terms of the bulge growth time $M_\bulge/\Md_\bulge$
  \BEQ
         V_\esc < V_\SNI \left (
                      \frac{\alpha_\SNI \epsilon_\SNI M_\bulge}
                           {t_\SNI \Md_\bulge}
                    \right )^{\frac{1}{2}} ~ .
  \EEQ
The above condition is plotted in Figure~\ref{blowout} (along with the
SNII condition) assuming the $\alpha_\SNI = \alpha_\SNII$.

  This oversimplified model does not readily produce a black hole blowout
condition, but it is possible to estimate the ratio of the black hole
mass to the bulge. If the black hole accretion is limited by the
Eddington luminosity, then $M_\BH = M_\BH^\init \exp[t/\epsilon_\BH
t_\edd]$, which sets the time scale for blowout $t_\blowout =
\epsilon_\BH t_\edd \ln [M_\BH/M_\BH^\init]$.  Combining the equation for
black hole luminosity with the blowout condition gives
  \BEQ
        M_\BH > \frac{t_\edd V_\esc^2}{\alpha_\BH c^2} \Md_\bulge
  \EEQ
If accretion onto the bulge is relatively constant, then $\Md_\bulge
\sim M_\bulge/t_\blowout$ and the ratio of the black hole to the bulge at
blowout is
  \BEQ
       \left (  \frac{M_\BH}{M_\bulge} \right )_\blowout
       > \frac{ (V_\esc/c)^2 }
              {\alpha_\BH \epsilon_\BH  \ln [M_\BH/M_\BH^\init]} ,
  \EEQ 
which for the characteristic numbers  $\epsilon_\BH  \ln [M_\BH/M_\BH^\init]
\sim 1$ and $V_\esc \sim 100~\kms$ requires a value of $\alpha_\BH >
10^{-4}$ in order to be in agreement with the observed  black hole to
bulge ratio of $\sim 10^{-3}$ (\cite{Faber96}).  Using this bound
for $\alpha_\BH$ we can now estimate the blowout condition
  \BEQ
         V_\esc <  \left (
                      \frac{c^2 \alpha_\BH}{t_\edd}
                      \frac{M_\bulge}{\Md_\bulge}
                    \right )^{\frac{1}{2}}
                   \left (  \frac{M_\BH}{M_\bulge}
                     \right )_\blowout^{\frac{1}{2}} ~ ,
  \EEQ
which is also shown in Figure~\ref{blowout} for $\alpha_\BH = 10^{-3}$.

\section{Discussion}
  The previous sections synthesized an inside-out galaxy formation
scenario by combining a spherical collapse model with the final
redistribution of gas by angular momentum. The aim was to interpolate
the gaseous evolution from models of the initial and final states of the
DM and the gas.

  The DM is governed by the specified final halo profile and $r_\max$
and $t_\collapse$, which are set by the mass overdensity of the initial
profile $\delta_m$.  The primary focus here is on the gas, and its
response to the DM potential.  Thus, a highly simplified model is adopted
of the post-collapse state of each DM halo.   Each DM shell stops
falling at the radius given by a pre-set final DM profile.  The validity
of this approximation is questionable.  Clearly it cannot properly
account for any hierarchical buildup of material in the halo.  If,
however, DM halos---in spite of hierarchical accretion---maintain a
quasi-self-similar density profile as is suggested from N-body
simulations (\cite{Navarro96}), then this asymptotic approach might be
OK.

  The state of the gas can be characterized by the evolution of the
adiabatic shock in response to the above DM potential.  For Gaussian
initial conditions, $r_\shock$ and $M_\shock$ can be related by simple
power law (Figure~\ref{shock_radius}) for a variety of DM halos.  As
mentioned earlier the adiabatic value of $r_\shock(M_\shock)$ is only an
upper limit.  In a real multi-phase gas, $r_\shock$ will be smaller due
to by the fraction of gas residing in cold clumps.  This effect will be
most pronounced earlier on, but during later stages of the infall the
adiabatic approximation is in good agreement with similar calculations
that take cooling into account (\cite{Thoul95}). 

  Using the power law for the shock evolution, the size of the initial
bulge (Figure~\ref{initial_bulge}) can be estimated by computing the
cooling time just inside the shock and setting it equal to the dynamical
time computed from the DM halo.  It is reassuring that the initial bulge
mass and radius are reasonable and display the correct dependence on the
halo parameters, except for the Burkert profile, which is only meant for
dwarf galaxy halos with $V_\circular \le 50~\kms$.  Unfortunately, the
sensitivity of these estimates to $\alpha_T$ and to the details of the
DM halo make this approach unsuitable for precise calculations. 

  After the infalling gas has been shocked, if it does not cool quickly
and form stars, then it will spiral in to its corresponding angular
momentum radius in about a dynamical time.  The simplest model for the
angular momentum of the gas is to assume that when each gas shell
reaches its maximum radius it is in approximately solid body rotation
with a frequency determined by $\lambda_\spin$.  What is most
interesting is the relative ease with which one gets an old bulge and a
young disk from this highly idealize picture.  The infall rates
(Figure~\ref{infall}), which are upper limits, are only a factor 10
greater than those seen observationally.  The size of the disk vs.  the
bulge depends upon the quantity $r_\bulge r_\halo \lambda_\spin$, thus
there is a large amount of freedom to create diverse objects.  More
complex angular momentum models (\cite{Ryden87}, \cite{Eisenstein95}),
which take into account the spectrum of initial fluctuations and
estimate the statistical distribution of angular momentum, could also be
used.  The overall effect of these models would be to shift and soften
the transition between bulge and disk infall. 

  The bulge will continue to grow slowly and may be halted if there is
sufficient energy input from supernova and/or a massive black hole to
drive a wind.  The disk will continue to grow outward as long as their
is infalling material to supply it.  The simple blowout estimates
computed in the previous section indicate that all the mechanisms
examined have the potential to drive a wind which may affect the
accretion onto the bulge.

\section{Conclusions and Further Work}

  A simple inside-out scenario for galaxy formation is presented, which
is a natural result of synthesizing earlier work on DM halos, spherical
collapse, and redistribution via angular momentum.  The scenario
predicts a straightforward sequence of events beginning with the
formation of an initial bulge due to the high densities and rapid
cooling times that exist when the first infalling mass shells are shock
heated.  After the initial collapse, the bulge continues to grow slowly
but may be halted by energy input from supernova and/or a massive black
hole initiates a wind that blows out the gas.  A disk forms as the
outermost shells collapse.  The higher virial temperatures and lower
gas densities of these shells result in longer cooling times; so that
instead of quickly forming stars the gas spirals in towards its
corresponding angular momentum radius.

  Although, this model is highly simplified and is not designed to
accurately describe the formation of any individual galaxy, it does (by
design) predict the overall features of galaxies.  In addition, it
appears that old bulges and young disks are an almost unavoidable 
result of this very simple model.  Furthermore the epochs at which
maximum infall into the bulge and inner disk take place are roughly
consistent with both the high star formation rates (\cite{Madau97}) and
quasar densities (\cite{Boyle97}) observed at $z \sim 2$.  A blowout
taking place early in the evolution of a galaxy is also consistent with
the observation of high velocity winds seen at high redshift
(\cite{Pettini97}).  This scenario also suggests two mechanisms for
preventing the formation of disk galaxies in the center parts of
clusters.  First, by encounters which disrupt the alignment of the
angular momentum vectors of the infalling gas. Second, by the high
virial temperatures present, which would cause late infall to be
absorbed into the ambient cluster gas. Of course, these are only two of
many possible explanations, but never-the-less they may be worth
exploring with more detailed simulations.

  These simple calculations suggest an idealized framework in which to
attempt the direct simulation of disks.  Cosmologically based
simulations are limited by the need to construct the potential from
particles, which leads to a significant increase in the complexity and
duration of the computations.  Observing 3D gas collapse into a
spherically symmetric time evolving potential like the one presented
here should be a useful test case of the 3D codes.  Further work in
this area should also focus on exploring the role of the processes that
intrinsically cannot be included in this simple model: hierarchical
structure formation and a multi-phase ISM.

  Finally, the extremely simple estimates of the blowout condition
suggest that Type I and  Type II Supernovae and super massive black
holes all have the potential to initiate an outflow and subsequently
change the evolution of the bulge.  Further calculations along these
line using more realistic assumptions could prove fruitful in
explaining the observed properties of bulges and elliptical galaxies. 


\acknowledgments

I would particularly like to thank my advisor Professor Spergel for his
important contributions to this work and many interesting and enjoyable
conversations.  I would also like to thank Professors Gunn, Ostriker,
Strauss and Gott for their helpful comments.  In addition, the comments
of an anonymous referee much improved this paper.  The author was
supported by NSF grant AST~93-15368. 


\newpage


\singlespace

\end{document}